\documentclass[twocolumn,showpacs,prl]{revtex4}

\ifx\pdfoutput\undefined
\usepackage{graphicx}
\else
\usepackage[pdftex]{graphicx}
\usepackage{epstopdf}
\fi

\usepackage[center]{subfigure}

\begin{document}

\title{An approximate hard sphere method for densely packed granular flows}

\author{Nicholas Guttenberg}
\affiliation{James Franck Institute}

\begin{abstract}

The simulation of granular media is usually done either with event-driven codes that treat collisions as instantaneous but have difficulty with very dense packings, or with molecular dynamics methods that approximate rigid grains using a stiff viscoelastic spring. There is a little-known method that combines several collision events into a single timestep in order to retain the instantaneous collisions of event-driven dynamics but also be able to handle dense packings. However, it is poorly characterized as to its regime of validity and failure modes. We present a modification of this method to reduce the introduction of overlap error, and test it using the problem of 2D granular Couette flow, a densely packed system that has been well-characterized by previous work. We find that this method can successfully replicate the results of previous work up to the point of jamming, and that it can do so a factor of 10 faster than comparable MD methods.

\end{abstract}

\pacs{45.70.Mg, 47.57.Gc, 02.70.Ns}
\maketitle

Granular material can take on a multitude of phases\cite{esipov1997granular} and regimes of behavior depending on the density, loading, and motion of the grains. Dilute granular flows in the absence of gravity behave like gasses initially, then form macroscopic structures in the form of vortices and finally clusters\cite{luding1999cluster}. Granular flows at higher densities exhibit a transition between liquid-like and solid-like behavior, eventually jamming\cite{corwin2005structural}. Accordingly, the demands on an algorithm for simulating granular flows vary depending on what regime of flow is being studied. 

For dilute granular gasses, the method of choice is event-driven hard-particle dynamics\cite{miller2004event, donev2005neighbor}, as the frequency of collisions between particles is low and so there are large savings in not having to simulate the intervening time. Furthermore, because the collisions are instantaneous it is possible to exactly conserve energy and momentum in the system (or to dissipate energy in such a way as to exactly satisfy the relationship for specific coefficients of restitution and friction). Even in the case of dilute granular gasses, however, it is possible for the system to undergo inelastic collapse, in which a cluster of particles form such that the time between collisions within the cluster approaches zero. Other methods exist that modify the event-driven method in order to continue past this point or to avoid this point (for instance the TC method\cite{luding1998handle} and a recent method that causes particles below a critical velocity to be treated as being asleep and frozen in place until they are woken by other collisions\cite{gonzalez2009extended}). These methods have their own advantages and disadvantages dependent on the specific nature of the problem. The TC method turns off inelasticity for repeat collisions within a short interval, modelling prolonged contacts of duration $t_c$ --- this prevents inelastic collapse of clusters of particles and has been shown to introduce controlled amounts of error for sufficiently short $t_c$, but slows down when the density becomes high due to the large number of collisions that occur in these short intervals. It is possible that a small cluster of particles can dominate the computational cost of simulating the entire system. The method of Gonzales et al., which was designed for grains piling up under gravity, does not suffer from a large number of collisions in the condensed phase, but requires a fixed rest frame for the grains. 

As the density of the granular flow increases, collapse-like situations become more common and the time between collisions can drop precipitously. As such, other methods are necessary to handle problems such as static granular piles, dense flows, and systems under compressive loading. The most commonly used method for dense flows is to use a discrete elements method that allows the grains to deform elastically\cite{ristow1992simulating, lee1993angle}. This involves introducing a potential associated with the overlap between grains and then iterating the equations of motion of particles under the given potential. This is not unreasonable as physical granular materials are not absolutely rigid. However, the Young's modulus of typical granular materials is quite high, from $~1$ (plastics) to $~100$ (copper, glass) GPa, which means that to simulate elastic grains with the same stiffness an extremely small timestep is necessary. Because of this, the computational grains are usually made many orders of magnitude softer than the physical grains. Additionally, the use of an interparticle potential with a timestepping method means that energy is no longer exactly conserved by the dynamics, either creating a long-term drift in the energy of the system for perfectly elastic simulations, or introducing error and velocity-dependence in the effective coefficient of restitution of the grains.

For nearly static packs, the soft-particle method can allow rearrangements that would not be physically possible for a hard-particle system. For these systems, the problem is usually modelled as a set of constraints determined by the existing contacts. For static systems, the constraint is that the net force on each grain must be zero. The constraints take the form of a linear system which can then be solved by matrix inversion\cite{tkachenko2000stress}. Depending on the contact network, the system of equations may end up having many non-trivial solutions or none. Also, the solution may contain spurious tensile forces, which necessitate a process of adaptively removing contacts under tensile stress\cite{tkachenko2000stress}. A similar method can be used for dynamical systems as well\cite{moreau1996numerical, unger2003contact}, but these methods are expensive (matrix inversion is $O(N^3)$) and can require iterative solution.

We present here a hybrid method based on work by McNamara, et al.\cite{mcnamara2000grains} that attempts to extend some of the benefits of hard-particle event-driven dynamics to dense flow regimes, while retaining the flexibility of soft-particle methods. The price we pay for this is that the method is no longer exact. The nature of the introduced error takes two forms: an inaccuracy in the effective grain sizes for grains that are moving fast relative to eachother, and a spurious sound speed akin to what is seen in soft methods that is approximately equal to the grain diameter divided by the timestep. The nature of these errors means that this method is appropriate in particular for dense flowing granular systems, where grains are moving locally in the same direction and there are few places where there is a static arrangement under load.

\section{Method}

The method is at its core an event-driven simulation. However, the collision prediction of the event-driven method is implemented using a fixed timestep. Now, rather than predicting the next event, we collect a list of events that will occur during the next timestep given the state of the system at the current timestep. Here we make a departure from the method of McNamara, et al.\cite{mcnamara2000grains}. In the original approach, collisions occured between grains that overlapped after they moved. A heuristic was used in which only approaching particles could collide, and only the first collision was dissipative. This was to prevent inelastic collapse, which would have the result of particles being permanently stuck in a state of overlap and thus creating spuriously high packing fractions. For dense systems, this can be disastrous, as the behavior of granular systems near jamming is extremely sensitive to minute changes in packing fraction.

Instead, we treat any pending overlaps that would occur between $t$ and $t+\Delta t$ as collisions taking place at time $t$. This way we prevent the initial overlap error from occuring. The collisions are handled using any instantaneous collision rules desired (frictionless, tangential restitution, frictional, etc). This update can create new collisions during the next timestep. One can then iterate the collision check until no collisions are found, but it is often more efficient to only iterate once and then correct for the overlaps introduced.

We also implement a correction scheme not present in the original method in order to correct for overlaps that do happen to occur for any reason. An example would be a poorly-prepared initial condition in which several grains began in an overlapping possition. When this occurs, we offset the grains along their contact normal to preserve their center of mass coordinate but place them just at the point of contact. Given two grains of radii and masses $r_1, m_1$ and $r_2, m_2$ a distance $d$ apart, the (outward) change in position of the grains is:

\begin{equation}
\delta_1 = m_2 \frac{r_1+r_2-d}{m_1+m_2} 
\end{equation}

\begin{equation}
\delta_2 = m_1 \frac{r_1+r_2-d}{m_1+m_2} 
\end{equation}

The error introduced by these adjustments occurs in two forms. The fact that we collide early means that particles moving quickly relative to eachother collide sooner than they would normally, and so they appear to have a slightly larger radius than they should. As the system cools (e.g. in a static pack, or for velocities in a dense flow becoming locally uniform), this error decreases, whereas in the original approach the error increases with cooling as particles are slower to leave an overlapped state than they are to enter it, and can therefore end up stuck. The second form of the error is the introduction of a spurious sound-speed controlled by the size of the timestep. This is because the collision check is performed only so many times per timestep, and so signals propagate at most that many grain diameters per timestep. From these error sources we can trade off more collision iterations for a larger timestep, or vice versa. 

The first form of error means that this method can experience slippage in static packs under load, as rather than actually being static the grains are in constant back and forth collision in order to transmit the load. The matrix methods mentioned above are more appropriate for such problems. Similarly, in a high relative velocity impact problem care must be taken with this method due to the high relative velocity between the grains. In flowing systems, however, the inherent fluctuations of the system tend to overcome the introduced errors. Relative velocities within the flow are usually much smaller than the flow speed, due to the effects of inelasticity, and so the grain size error is small. 

\section{Validation}

We wish to validate our modified method against a system at the boundary between flow and solidity, as this is where we expect problems to crop up and also where we can find  existing results from other methods operating in their regimes of validity. The problem we choose is two-dimensional granular couette flow\cite{howell1999stress, latzel2003comparing, campbell2006computer}.

In comparing our simulation with experimental and numerical realizations of granular Couette flow, we focus on two flow transitions. First, we show that we can reproduce the strengthening transition that has been observed\cite{howell1999stress} near the packing fraction $\phi=0.776$. This is the point at which the grains are in sufficiently persistent frictional contact to maintain a distribution of stress across the entire system. As this is primarily a flowing system case without jamming, we expect our code to do well here.

We then turn to the transition between shear flow and blocked flow around the critical packing fraction near $\phi=0.8$ (where two-dimensional random close pack is $\phi=0.82$). As we approach the transition, we are moving from a flowing granular system to a jammed granular system under load. This is an extreme test of our algorithm, as the overlap method could in principle introduce spurious rearrangements that break the jamming of the system and allow for flow where there should not be flow. As such we can use this system to see how the algorithm begins to have difficulties, and what sort of errors will crop up.

The setup of our simulation consists of a set of disc-shaped grains of uniform density with Coulomb friction coefficient $0.5$ and normal restitution coefficient $0.9$. The grains are polydisperse, being uniformly distributed in radius between $0.8$ and $1.2$.
The grains are packed into the region between an outer cylinder of radius $90 \langle R \rangle$ and an inner cylinder of radius $30 \langle R \rangle$ (where $\langle R \rangle = 1$ is the average grain radius). In order to apply the boundary conditions treat any collision of a grain with the boundaries as if that grain had collided with another infinitely massive grain. The cylinders have the same coefficient of restitution and coefficient of friction as the grains. The inner cylinder is rotated at an angular velocity $\Omega=1$ and the outer cylinder is held fixed. 

We determine the initial pack by evaluating the number of grains we expect to fit in the area between the cylinders and attempting to place that many grains without overlap. When overlap occurs (as it will for sufficiently high packing fractions) we relax the position of the grains in the system to minimize the total overlap. Due to the distribution of grain sizes, the actual packing fraction we achieve will vary by about 1\% from the packing fraction we attempt to create. We measure and record the actual packing fraction and use that number for all of our plots and comparisons. Over the range of packing fractions we explore our system contains between $5040$ grains ($\phi = 0.707$) and $6264$ grains ($\phi = 0.878$).

\subsection{Stress Profile}

\begin{figure}[t]
\includegraphics[angle=0]{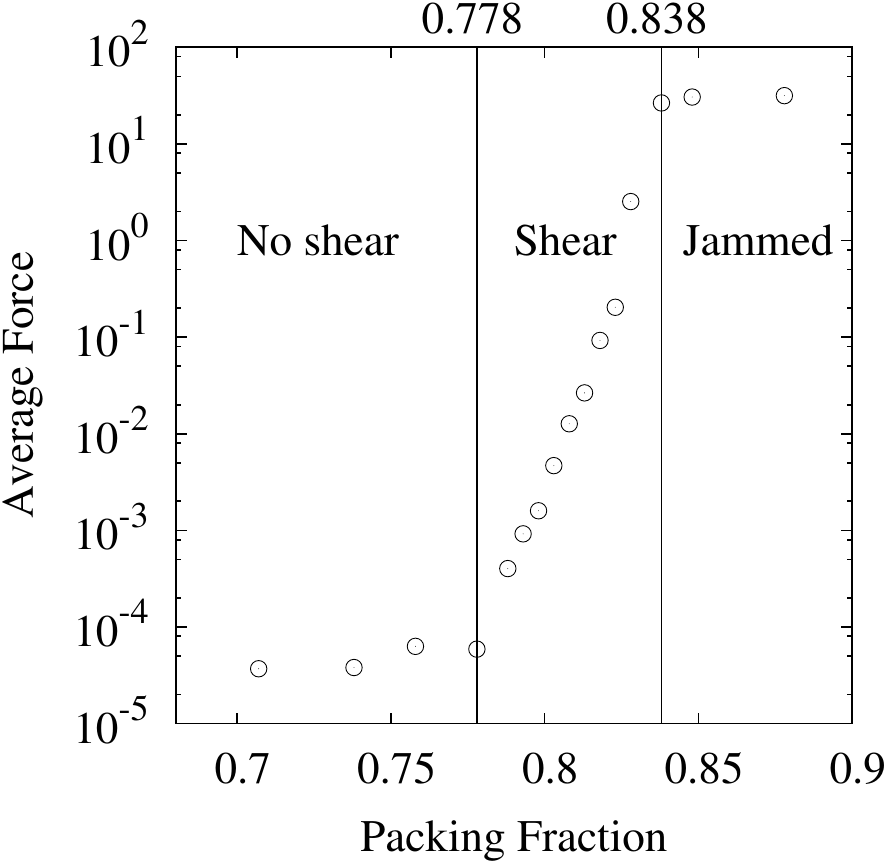}
\caption{The proxy stress as a function of packing fraction for simulations using timestep $10^{-5}$. The strengthening transition can be seen at $\phi=0.778$, and the transition to jammed flow at $\phi=0.838$. }
\label{StressProfile}
\end{figure}

First we check whether or not we can observe the strengthening and jamming transitions with our code, and how our code handles the onset of jamming. We do this by examining the stresses in the system, which should be zero in the case where there is no communication between the center wall and outer wall due to the formation of a gap, begin increasing after the strengthening transition, and then saturate when the system jams. Because we have instantaneous collisions rather than persistent contacts, we will measure an analog of the stress that does not exactly correspond in the jammed limit. Specifically, we measure a proxy of the stress on grains by averaging the magnitude of normal momentum transfer and tangential momentum transfer (that is to say, the impulses) over a period of simulation time. This gives us a measurement of the average force experienced by a grain. 

The average stress tensor for a granular system can be determined\cite{radjai1996force}:

\begin{equation}
\sigma_{ij} = \rho \langle F_i d_j \rangle
\end{equation}

Here $\rho$ is the number of contacts per unit volume, $F_i$ is the i component of the force on the grain, and $d_j$ is the j component of the vector corresponding to the contact. If we take this as emerging from a series of collisions, then we can add up the number of events, replace the forces with impulses, and divide by the interval. As such we can relate our proxy stress $S$ to the stress tensor:

\begin{equation}
S = \frac{\sum I^{n}_i d^{n}_i + I^{n}_j d^{n}_j }{\tau} = \pi r^2 (\sigma_{xx} + \sigma_{yy})
\end{equation}

Therefore in the case of a time-averaged collection of instantaneous collisions, this proxy stress corresponds to the trace of the stress tensor (as measured for granular systems in \cite{radjai1996force} and others). However, in the case of sustained contacts, in our method the number of collisions depends on the timestep, and so the correspondance is less clear. In practice if the bounding force is of fixed magnitude, then each collision imparts some fraction of that momentum flux, and so the quantity we measure would still represent the trace of the stress tensor, simply divided up in an arbitrary number of collisional chunks. However, if the bounding force can become infinite (due to, say, the forced addition of a particle to a jammed pack), then this quantity will not similarly diverge to infinity but will saturate at a maximum based on the number of collisions per timestep. When we explore packings beyond jammed using this method, the stress resisting motion should become infinite, but we instead observe a saturation at a large finite value due to the nature of the algorithm. 

The stress profile we observe when using our smallest timestep ($10^{-5}$) is shown in Fig.~\ref{StressProfile}. We can immediately notice the strengthening and jamming transitions. The strengthening transition first shows up at the data point at $\phi=0.778$, which is consistent with the $0.776$ result of \cite{howell1999stress}. The jamming transition on the other hand behaves somewhat differently. In previous work it is observed that the flow locks up near $\phi=0.8$, but we can observe shear flow out to $\phi=0.828$, slightly beyond the value of random close pack. As we increase the packing fraction, the system locks up near the outer ring and this locked region spreads inwards. By $\phi=0.838$ the system is observed to have locked up entirely and is undergoing rigid body motion rather than shear flow. At this point we observe slippage at the walls and unsteady drifting rigid body motion of the granular pack. 

\begin{figure}
\includegraphics[angle=0,width=0.9\columnwidth]{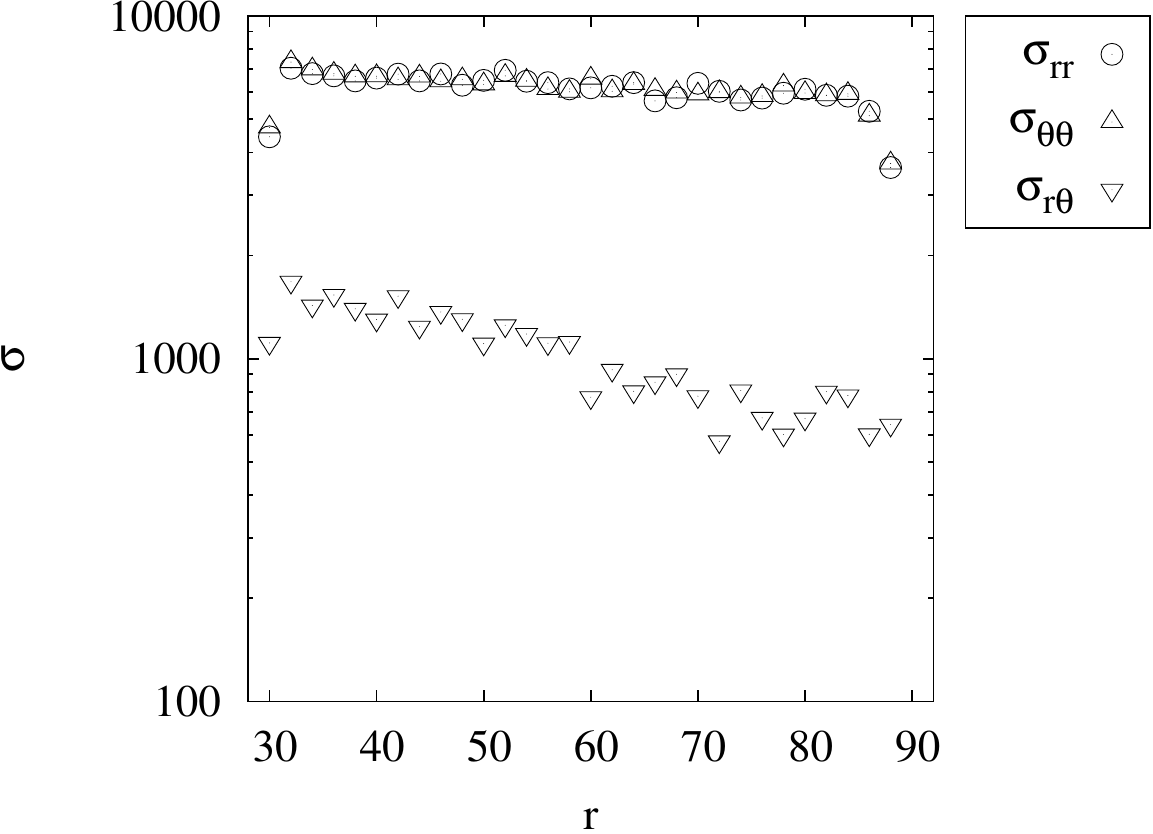}
\caption{This figure shows the separate components of the stress tensor measured in a simulation at $\phi=0.828$.}
\label{StressComponents}
\end{figure}

\begin{figure}
\includegraphics[angle=0,width=0.9\columnwidth]{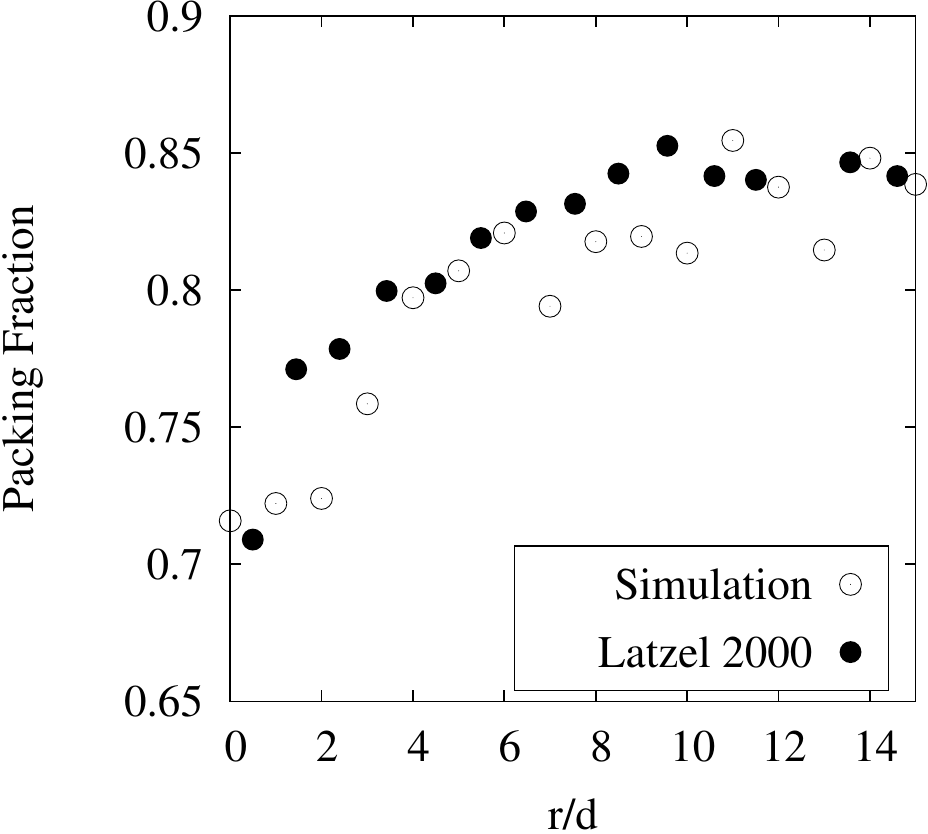}
\caption{This figure shows the local packing fraction as a function of distance from the inner ring in our simulation (hollow points). We compare with data from \cite{latzel2000macroscopic} (solid points). There is a local decrease in packing fraction near the inner ring, corresponding to the formation of a shear band.}
\label{PackingFraction}
\end{figure}

In granular cylindrical couette flow, a shear band forms near the inner moving cylinder. This manifests as a local reduction in the packing fraction compared to the global packing fraction. The packing fractions $\phi$ we report are the global values of the packing fraction, and so the system will be packed somewhat more tightly away from the inner ring and somewhat more loosely near the inner ring. We compare the packing fraction distribution near the inner ring with with a similar setup from \cite{latzel2000macroscopic} in Fig.~\ref{PackingFraction}. We note that we have a separation of $30$ grain diameters between our inner and outer ring, whereas theirs is a $20$ grain diameter separation, and so we only compare the section of the curve near the inner ring. We also show the spatial distribution of stresses in a single simulation at $\phi=0.828$ for each component of the stress tensor measured in the radial/azimuthal basis in Fig.~\ref{StressComponents}. 

This jamming of the system corresponds to a saturation of the proxy stress in the stress profile. Furthermore, it is observed that the value of the proxy stress is timestep dependent, even though the observed transition points are not. Let us try to understand this.

When the system is jammed, the proxy stress should approach the force used to drive the center cylinder. However, in our simulation the motion of the center cylinder is specified and so this limiting stress is infinite. We can never observe an infinite force because we do not have persistent contacts. Instead, each collision will transfer the same amount of momentum (since they are basically transporting momentum between the two boundaries without retaining any locally). As such, the proxy stress we expect to measure should be proportional to the number of collisions per unit time, and so the limiting value of the stress in the jammed case is in fact expected to be timestep dependent. This is a consequence of the way we measure the stress rather than a sign of timestep dependent physics of the system. As such, we should look for the qualitative features of the stress profile rather than expect a quantitative comparison with any given soft-grain system. In essence, the algorithm can capture the steadiness of the jammed regions of a flow over intermediate timescales, but has problems quantitatively reproducing the forces inside the pack as there is no independent degree of freedom associated with those forces. In order to make quantitative comparisons and to check the convergence of the algorithm with respect to timestep size, we must make use of a different observable. As such we turn to the velocity profiles of the simulated flows.

\subsection{Velocity Profile}

\begin{figure}[t]
\includegraphics[angle=0]{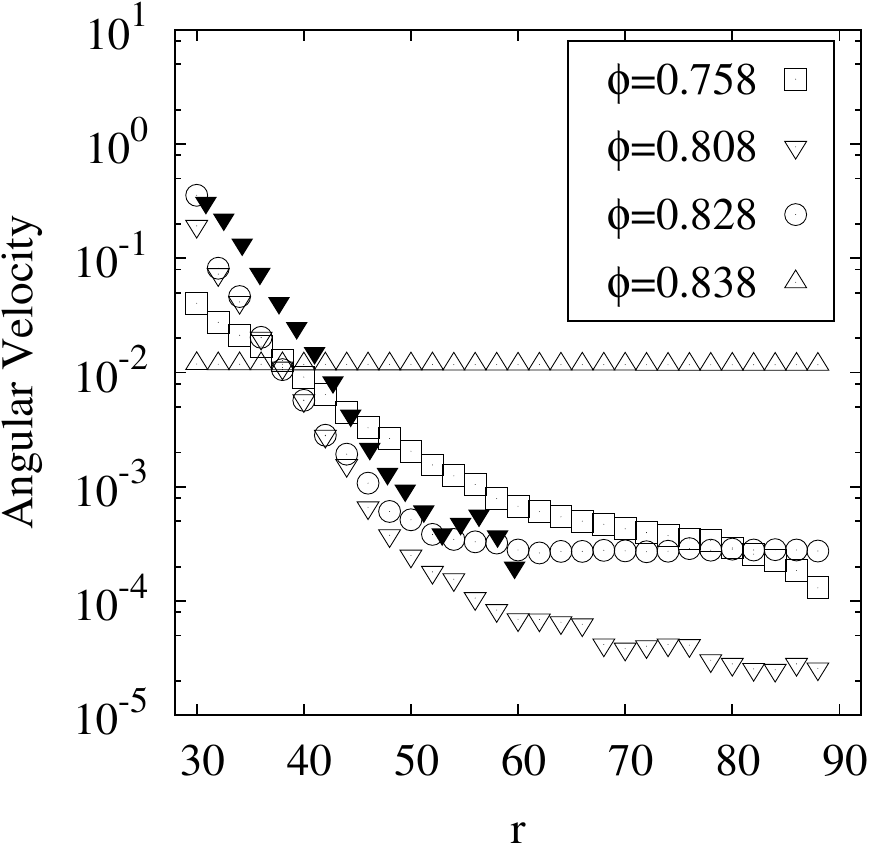}
\caption{This figure shows the measured angular velocity profiles for different packing fractions. For packing fractions below the strengthening transition, a broad exponential decay is observed. Beyond the strengthening transition, the exponential decay becomes significantly sharper, corresponding to a shear band of a few grain diameters. At sufficiently high packing fractions, parts of the system begin to jam and undergo rigid body rotation around the inner cylinder, eventually filling the system with a jammed state. The solid points are velocity profile data at $\phi=0.804$ from\cite{latzel2003comparing}. The bump between $r=50$ and $r=60$ in the solid data is consistent with the scale of fluctuations between different data sets in that paper.}
\label{VelocityProfiles}
\end{figure}

\begin{figure}[t]
\includegraphics[angle=0]{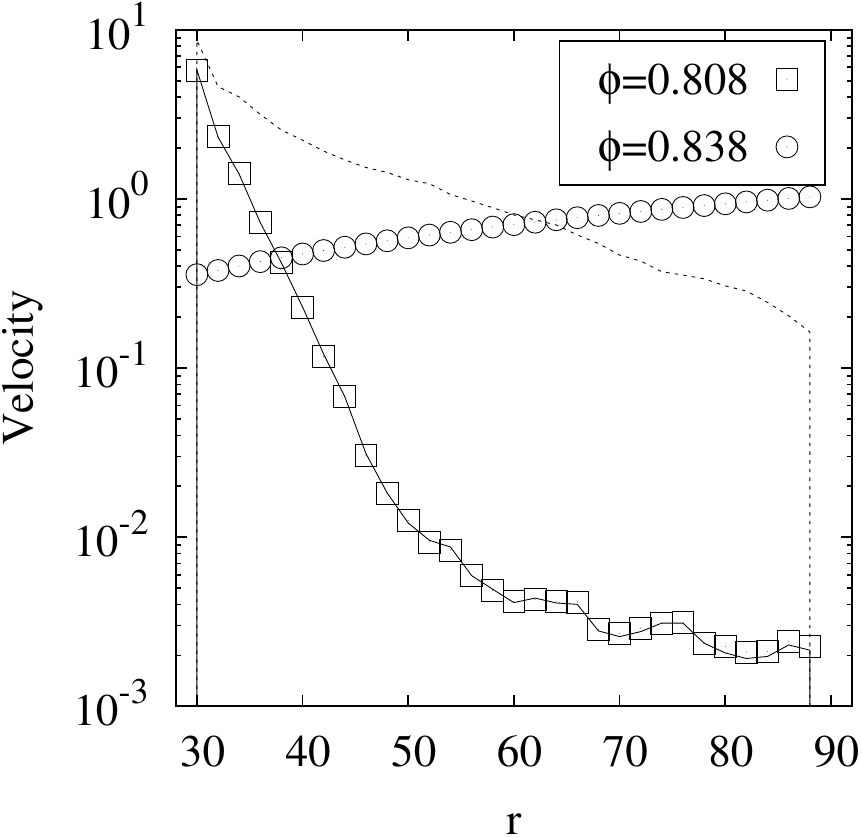}
\caption{This figure shows the measured (points) versus internal (lines) velocity profiles for two packing fractions. Below the jamming transition, the internal velocity variables agree with the measured average displacement of the grains. Once the jamming transition occurs, the overlap correction algorithm causes the internal velocity variables to disagree with the average displacement.}
\label{VelocityComparison}
\end{figure}

We now look at the velocity profiles observed in our simulation. Again we use our smallest timestep $10^{-5}$, and plot angular velocity profiles corresponding to different packing fractions (Fig.~\ref{VelocityProfiles}). In this plot, rigid body motion corresponds to a flat line. For packing fractions that do not support shear, the velocity profile exhibits a broad exponential decay with a lengthscale of $5$ grain diameters. Once the packing fraction passes the strengthening transition, the exponential decay becomes faster, and a shear band with a decay length of $1.2$ grain diameters is observed to emerge. In this range we have plotted data from \cite{latzel2003comparing} against our velocity profiles they have the same initial exponential character followed by a levelling off at the same point. Beyond this regime, at the onset of jamming, the angular velocity of the entire granular ensemble becomes constant, and the grains between the two cylinders begin to undergo slow rigid body rotation around the inner cylinder.

We make one cautious note about the measurement of these velocity profiles. Each grain in the simulation has internal velocity variables, and the velocity can normally be read out by examining the averages of these values. However, the actual net displacement that a grain experiences may be modified slightly by the method used to offset grains so that they are no longer overlapping. Within the regimes of free flow and shear flow, the internal velocity is observed to be the same as the velocity one gets by directly measuring the displacement of grains with respect to time. However, once the grains have jammed, this changes. In a jammed pack, the redistribution of momentum following a collision will without fail introduce a new collision, as there is no where for a grain to go beyond following the mean motion of its neighbors. The various momentum transfer mechanisms (collision, restitution, and friction) can cause the velocity of a grain to converge to that of its neighbors in the bulk, but there is always some tension between the boundary condition at the center and that at the exterior, which each drive the mean velocity of the granular ensemble to a different value. As such, the internal velocities can never completely relax, and overlaps are generated at every timestep. The overlap correction step then adjusts for this, moving the particles around the system in ways not accounted for by their internally recorded velocities. This maintains the jammed state and prevents rearrangements successfully, but causes the internal velocities to be inaccurate representations of the actual grain movement. We see this in Fig.~\ref{VelocityComparison}, where we compare the directly measured velocities (points) with the internal velocities (lines) for an unjammed and jammed state.

We have now seen that our code behaves reasonably for packing fractions that sustain shear. When the system is made to jam, the algorithm relies strongly on the overlap correction step to retain the rigid nature of the system, but as a consequence fails to capture the forces and velocities of the jammed particles in a timestep independent way. Despite these issues, the algorithm fails gracefully, maintaining the jammed pack without rearrangements for packing fractions only slightly into the jamming phase. This suggests that the algorithm should be appropriate for establishing the proper immobilized regions in a larger flow, since the situation is not one with frustrated overlaps (overlaps can always escape to the boundary of the immobilized regions and be absorbed in density fluctuations in the exterior flow).

\subsection{Effects of timestep size}

\begin{figure}[t]
\includegraphics[angle=0]{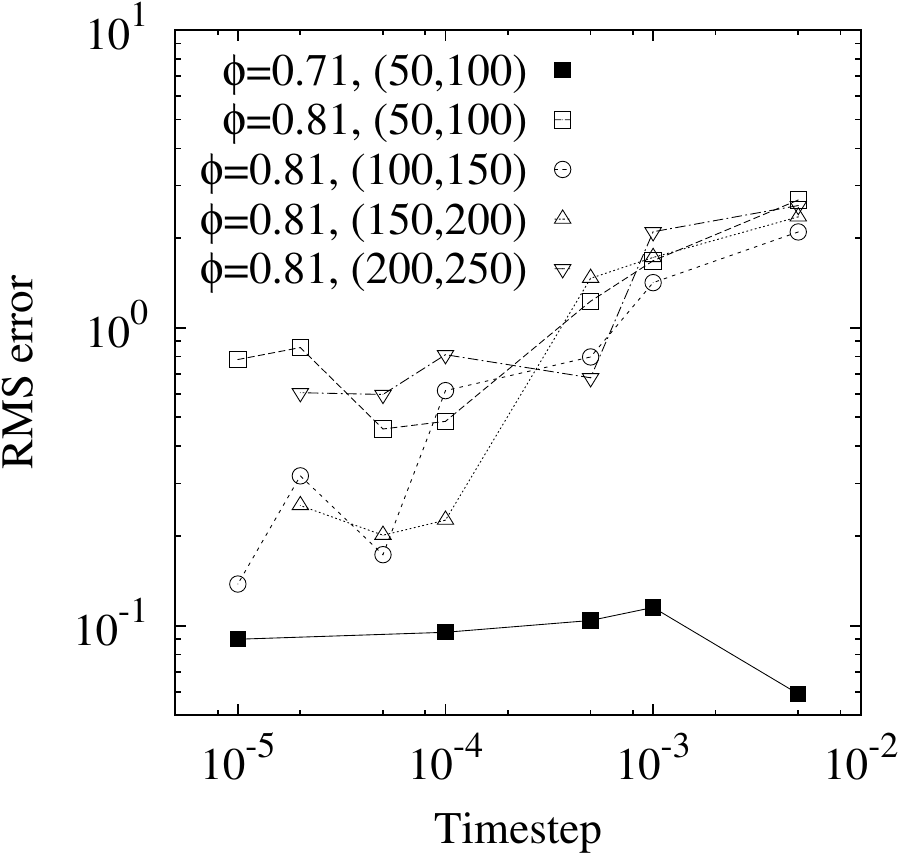}
\caption{This figure shows the convergence of the algorithm for two packing fractions ($0.71$ and $0.81$) during several different intervals of physical time ($t=50..100,\ 100..150,\ 150..200,\ 200..250$). The algorithm performs with very little error in the less dense packing --- for all timesteps shown, the error appears to be below the noise floor. In the case of the denser packing, systematic differences from the noise floor emerge for timesteps larger than $10^{-4}$. }
\label{FigConvergence}
\end{figure}

The algorithm presented in this paper is an approximation of the hard sphere limit. As such, it is useful to estimate the error introduced by the approximation. We do this by examining the timestep dependence of the mean velocity profile. This procedure reveals something of possible concern - the velocity profile converges to its steady state value apparently logarithmically with time, corresponding to infrequent opportunities to rearrange the unmoving region of the system. Due to this the simulation time needed to compare very large timesteps and very small timesteps out at the steady state is prohibitive. Instead, we choose a fixed physical time interval over which to compare the velocity profiles. For larger timesteps, this may introduce additional error as the number of independent samples taken of the velocity profile is inversely proportional to the timestep. The signature of this would correspond to a $\sqrt{\Delta t}$ contribution to the error.

We measure the velocity profile during a fixed time interval. We then compute the root mean square error between the profile observed at the trial timestep and the profile calculated using a timestep of $5\times 10^{-6}$. The convergence of this error is plotted in Fig.~\ref{FigConvergence} for a packing fractions of $0.71$ and $0.81$ for different physical intervals. The algorithmic convergence is not a simple monotonic progression of the error to zero. Instead, in the case of $\phi=0.81$ it decreases initially (for large timesteps) and then levels out into a non-monotonic scatter of points. In the case of $\phi=0.71$ it does not decrease at all, though the magnitude of the error is lower in all cases. 

We can understand this from the fact that the individual grain dynamics are quite chaotic, and that the structure of the jammed portion of the pack near the edge of the shear band can have a strong effect on the rate of momentum transfer and is also glassy in its structure and therefore does not rearrange easily. This means that there are large fluctuations in the instantaneous velocity profile even at long times, and that even for the same initial condition, the fluctuations will quickly become decorrelated in the presence of any error whatsoever. The non-monotonicity and wide variation across different physical time intervals of the error at very small timesteps (less than $10^{-4}$) suggests that the measured error here is being dominated by the noise floor introduced by these fluctuations. In the case of the smaller packing fraction $\phi=0.71$, the error appears to be independent of timestep size, suggesting that the system is converged even for the largest timestep shown here. 

\section{Conclusions}

We have presented a method for simulating dense, mobile granular flows using a hybrid of hard-sphere and molecular dynamics approaches extended from the method of \cite{mcnamara2000grains}. The resultant method can be efficiently and easily implemented and does not suffer from inelastic collapse. Furthermore, we have shown that the algorithm remains functional even into jammed states with persistent contacts, so long as the timestep used is sufficiently small so that unphysical rearrangements do not occur during the period of simulation. The algorithm successfully reproduces the packing fraction value at which the strengthening transition in granular Couette flow occurs and produces velocity profiles consistent with those measured in other simulation work\cite{latzel2003comparing}. 

Furthermore, the algorithm converges up to the noise floor introduced by long-time fluctuations at large timesteps ($> 5\times 10^{-3}$) when the packing fraction is small, and continues to converge even as the packing fraction approaches the jamming point for sufficiently small timesteps ($< 10^{-4}$). These timesteps are significantly larger than what would be needed to properly resolve the extremely stiff 'soft' particles as normally used in MD-based simulations of granular material (for example, a timestep of $10^{-5}$ was used in\cite{buchholtz1998interaction} for a fairly dilute stream impacting an obstacle). We therefore suggest that this algorithm is well-suited to dense but flowing granular systems and can perform as much as a factor of ten faster than other existing methods.

\section{Acknowledgements}

We would like to acknowledge Wendy Zhang, Heinrich Jaeger, and Sid Nagel for helpful discussions and suggestions as to good validations for this algorithm. We would also like to acknowledge Thorsten P{\" o}schel for his insights and comments on this algorithm, which motivated several improvements. This work was funded by a University of Chicago NSF MRSEC Kadanoff-Rice Fellowship.

\bibliographystyle{apsrev}

\bibliography{bibliography}

\end{document}